\documentclass[apj]{emulateapj}
\usepackage{epsfig}
\usepackage{amssymb}
\usepackage{amsmath}
\usepackage{graphicx}
\usepackage{latexsym}
\usepackage{dcolumn}
\usepackage{bm}
\usepackage{ulem}
\usepackage{multirow}

\def\gsim{\;\rlap{\lower 2.5pt
 \hbox{$\sim$}}\raise 1.5pt\hbox{$>$}\;}
\def\lsim{\;\rlap{\lower 2.5pt
   \hbox{$\sim$}}\raise 1.5pt\hbox{$<$}\;}
\DeclareMathOperator*{\argmin}{arg\,min}
\def\ie{{\it i.e. }}
\def\eg{{\it e.g. }}
\def\dd{\,\mathrm{d}}
\def\bfx{\mathbf{x}}
\def\bfy{\mathbf{y}}
\def\bfk{\mathbf{k}}
\def\bfu{\mathbf{u}}
\def\bfr{\mathbf{r}}

\shorttitle{Removing the Impact of Correlated PSF Uncertainties in Weak Lensing}
\shortauthors{Lu et al.}

\begin{document}
\title{Removing the Impact of Correlated PSF Uncertainties in Weak Lensing}

\author{Tianhuan Lu$^{1,2}$, Jun Zhang$^{1,3*}$, Fuyu Dong$^1$, Yingke Li$^1$, Dezi Liu$^4$, Liping Fu$^5$, Guoliang Li$^6$, Zuhui Fan$^4$}
\affil{
$^1$Department of Physics and Astronomy, Shanghai Jiao Tong University, Shanghai 200240, China \\
$^2$Department of Physics, UC Santa Barbara, Santa Barbara, CA, 93106, USA \\
$^3$Shanghai Key Laboratory for Particle Physics and Cosmology, Shanghai 200240, China \\
$^4$Department of Astronomy, School of Physics, Peking University, Beijing 100871, China\\
$^5$The Shanghai Key Lab for Astrophysics, Shanghai Normal University, Shanghai  200234, China\\
$^6$Purple Mountain Observatory, Chinese Academy of Sciences, Nanjing 210000, China
}

\email{*betajzhang@sjtu.edu.cn}

\begin{abstract}

Accurate reconstruction of the spatial distributions of the Point Spread Function (PSF) is crucial for high precision cosmic shear measurements. Nevertheless, current methods are not good at recovering the PSF fluctuations of high spatial frequencies. In general, the residual PSF fluctuations are spatially correlated, therefore can significantly contaminate the correlation functions of the weak lensing signals. We propose a method to correct for this contamination statistically, without any assumptions on the PSF and galaxy morphologies or their spatial distribution. We demonstrate our idea with the data from the W2 field of CFHTLenS. 

\end{abstract}

\keywords{cosmology, large scale structure, gravitational lensing - methods, data analysis - techniques, image processing}

\section{Introduction}
\label{sec:intro}

Weak lensing refers to the systematic shape distortions of the backbround galaxy images by the foregound density fluctuations. It provides a direct probe of the cosmic large scale structure (\cite{bs2001}; \cite{refregier2003}; \cite{hj2008}; \cite{kilbinger2015}). The most well known source of systematic errors in weak lensing measurements is due to the presence of the point spread function (PSF). Correction for the PSF effect has been extensively discussed within various shear measurement methods (\cite{kaiser1995}; \cite{lk1997}; \cite{hoekstra1998}; \cite{kaiser2000}; \cite{rhodes2000}; \cite{bj2002}; \cite{hs2003}; \cite{rb2003}; \cite{mr2005}; \cite{kuijken2006}; \cite{miller2007}; \cite{nb2007}; \cite{kitching2008}; \cite{zhang2008}; \cite{ba2014}; \cite{zhang2015}; \cite{bernstein2016}). An important issue is about the reconstruction of the PSF at the galaxy positions. The residual PSF errors can affect the accuracy of shear measurement in terms of shear averages and shear-shear correlations. \cite{berge2012} compares the residual correlations of PSF ellipticities for three common interpolations -- polynomial, Delaunay triangulation, and Kriging -- and points out that Kriging interpolation is the best choice based on simulations. \cite{chang2012} proposes PSFENT and compares its ellipticity residual correlation to other interpolations and shows that PSFENT has a great advantage over the others. \cite{lu2017} (L17 hereafter) compares residual correlations in weak lensing measurements for four sets of methods -- global polynomial, chipwise polynomial, Kriging, and Shepard -- using the real data from CFHTLenS, and claims that 1st-order chipwise polynomial performs the best. 

The spatial variations of the PSF exist on a wide range of scales. Accurate weak lensing measurements require accurate modelling and correction of such variations on all of the scales. Although large scale variations ($\gsim 10\,\mathrm{arcmin}$) can be reconstructed relatively accurately using PSF reconstruction methods (polynomial fitting, Kriging, etc.), it is impossible to reconstruct small scale variations ($\lsim 10\,\mathrm{arcmin}$) on the object-by-object basis due to the limitation of star number density and the stochasticity caused by atomspheric turbulence. As is pointed out by \cite{heymans2012}, high frequency variation has a great impact on weak lensing measurements on scales $\lsim 10\,\mathrm{arcmin}$ in terms of correlation functions. Testing this impact using real data from CFHTLenS, L17 find that the influence of residual correlation is comparable to typical shear-shear correlations. Even though measuring cross-correlation between different exposures can mitigate the problem, the residual correlations are still statistically significant. 

The problem of the presence of residual correlations is of great importance in measuring shear-shear correlations due to the following reasons: 1) it is a systematic error, which will not be reduced by adding data, and will be more significant when the statistical error is lowered; 2) the effect of residual correlations on shear correlations is appreciable (up to 10\% according to L17); and 3) its effect on shear correlation also depends on the morphology of galaxies (\eg their average size) so that it cannot be easily calibrated. Existing methods \citep{rowe2010,jarvis2016,zuntz2017} use the correlation of the residual PSF ellipticities and sizes, as well as the distribution of the model PSF parameters and galaxy sizes, to estimate and correct for the residual PSF effect on the shear-shear correlation. A key assumption in these studies is the formula relating the PSF error to the bias of the cosmic shear \citep{paulin2008}. Nevertheless, the accuracy of this formula is questionable for concurrent practical shear measurement methods. For example, the shear bias may not only depend on the variations of the PSF size and ellipticities, but also the fluctuations of the higher order PSF shape moments. Similarly, the high order morphological parameters of galaxies likely play an important role through their couplings to the PSF moments as well. It would be ideal to have a method that can correct for the fluctuations of the PSF uncertainties within a specific and practical shear measurement method, without assumptions on the morphologies of the galaxies or the PSF. This is the motivation of our work.
   
For our purpose, we adopt the shear measurement method described in Zhang et al.(2015) (ZLF15 hereafter). It has the advantage of being free of shear biases in the absence of PSF uncertainties, and free of assumptions about the galaxy or PSF morphology (L17). By parameterizing the morphologies of the residual PSF images through the Principal Component Analysis (PCA), we find that the shear biases have very good linear relations with the coefficients of the principal components (called PCs hereafter) of the residual PSF image. The PCA description of the PSF residuals is more accurate and complete than that based only on the ellipticity components and the PSF size. To remove the bias in the shear-shear correlation due to the spatially correlated PSF residuals, one can introduce additional PSF components with anticorrelated PCs. In \S\ref{sec:method}, we describe the details of our method. In \S\ref{sec:results}, we present the performance of our method on CFHTLenS data. Finally, we conclude in \S\ref{sec:conclusion}.


\section{Method}
\label{sec:method}

\subsection{Overview}

In L17, the additive and multiplicative biases $\{c_1, c_2, m_1, m_2\}$ are used to quantify the effect of the residual PSF error on the shear measurement. They are defined as: 
\begin{eqnarray}
\tilde{g}_1&=&(1+m_1) g_1+c_1,\nonumber \\
\tilde{g}_2&=&(1+m_2) g_2+c_2,
\end{eqnarray}
where $\{g_1,g_2\}$ denote the true reduced shears, and $\{\tilde{g}_1,\tilde{g}_2\}$ denote the measured reduced shears that are evaluated with the shear measurement method of ZLF15 at the positions of the reference stars. At the location of each reference star, a large number of mock galaxy images of random morphologies are generated, sheared, and convolved with the PSF given by the reference star. The shear signal is recovered with the PSF model constructed from another group of stars in the same field that are responsible for PSF reconstruction. The resulting shear biases ($c_1,c_2,m_1,m_2$) calculated at every reference-star position directly represent the effect of residual PSF uncertainties on shear measurement over the field of view. Its impact on shear statistics such as shear-shear correlations can be quantified straightforwardly thereafter. Note that the PSF model in this paper always refers to the power spectrum of the PSF image, which is the only PSF information required by ZLF15 for shear recovery. 

It is shown in L17 that the impact of residual PSF error on statistics such as shear averages is not significant at all, due to the cancellation of the errors on a large galaxy sample. Its impact on the shear-shear correlations is however not negligible. It is mainly characterized by the spatial correlations of the additive biases. To solve this problem, our strategy is to connect the additive bias with the PSF residual.

In our method, we use PCA to model PSF residuals so that each residual image can be characterized by a few parameters. PCA is performed on the residual images $\bfr(\bfk)$ of the PSFs, which is defined as pixel by pixel differences between the reconstructed PSF and the true PSF (star), to get a set of principal coefficients $r_j$ for each residual image. Then, we assume that the additive shear biases $c_i$ are approximately linearly related to these principal coefficients $r_j$ of the residual PSF images, with higher order terms ignorable, \ie 
\begin{equation}
c_i=\sum_{j}{\alpha_{ij}r_j}+\mathrm{O}(r_j^2)
\label{eqn:bias-expansion}
\end{equation}
The values of $\alpha_{ij}$ depend on the galaxy and PSF morphology. For example, larger galaxies usually correspond to smaller $\alpha_{ij}$, meaning that the shear recovery accuracy is less sensitive to the PSF reconstruction uncertainty. The linear relationship remains accurate as long as the coefficients $r_j$ are small. 

Eqn.\eqref{eqn:bias-expansion} tells us that the contamination to the shear-shear correlations are contributed by the spatial correlations of the PC coefficients of the residual PSF. The later can be studied using only the star samples. These facts motivate us to correct for the impact of PSF residuals on the shear-shear correlations through the following two steps: 

\begin{enumerate}
\item For a given PSF reconstruction method, we first get the distribution of the residuals of the PSF power spectra, and then perform the Principal Component Analysis on the power spectrum images to identify $n$ PCs that can significantly affect shear recovery. This is followed by the measurement of the spatial correlations $\xi_{pq}$ ($p,q=1,2,...,n$) of the PC coefficients at different locations. The details are given in \S\ref{sec:res-field}.

\item The two-point correlation functions $\xi_{pq}$ are used to generate $n^2$ Gaussian random fields $w_{pq}(\bfx)$ ($p,q=1,2,...,n$). We then build two separate PSF fields by adding the products of $w_{pq}(\bfx)$ and the corresponding PCs to the PSF model in two different ways. The two PSF fields lead to two shear catalogs, the cross-correlation of which yields a shear-shear correlation measurement that is free of systematic biases from the PSF uncertainties. This is shown in \S\ref{sec:corr-field}.
\end{enumerate}


\subsection{Characterizing the PSF Uncertainties with PCA}
\label{sec:res-field}

We choose the 1st-order chipwise polynomial fitting method to model the PSF distribution, because it is found to perform the best in L17. Note that we interpolate the power spectra of the PSFs, \ie star images in the Fourier space, so that the images are automatically centered. The power spectrum is the only PSF information required by ZLF15 for shear recovery. The interpolation is done pixel-by-pixel in Fourier space.

A residual (spectrum) image $\bfr(\bfk)$ is defined as: 
\begin{equation}
\bfr(\bfk):=\bfu(\bfk)-\bfu_0(\bfk),
\label{eqn:res-def}
\end{equation}
where $\bfu(\bfk)$ is the modelled and normalized PSF spectrum, and $\bfu_0(\bfk)$ is the true one (from a star image). Note that the residual images can only be calculated at the star positions, not the galaxy positions. In practice, we randomly divide all star images into three groups: the ``reconstruction'' group (1/2 of all stars), the ``reference'' group (1/4 of all stars), and the ``validation'' group (1/4 of all stars). The stars in these three groups are used in the following ways: 

\begin{itemize}
\item The reconstruction group: The stars are used to reconstruct the PSF power $\bfu(\bfk)$ at any position. 
\item The reference group: The stars are treated as true PSFs $\bfu_0(\bfk)$. They are compared with corresponding reconstructed PSFs to get residuals and residual correlations. 
\item The validation group: The stars are reserved as galaxy PSFs, and are never shown to our removal method. The removal method will be asked to reconstruct PSFs at validation star positions so that we can test its performance.
\end{itemize}

\begin{figure}[!htb]
  \centering
  \includegraphics[width=8cm]{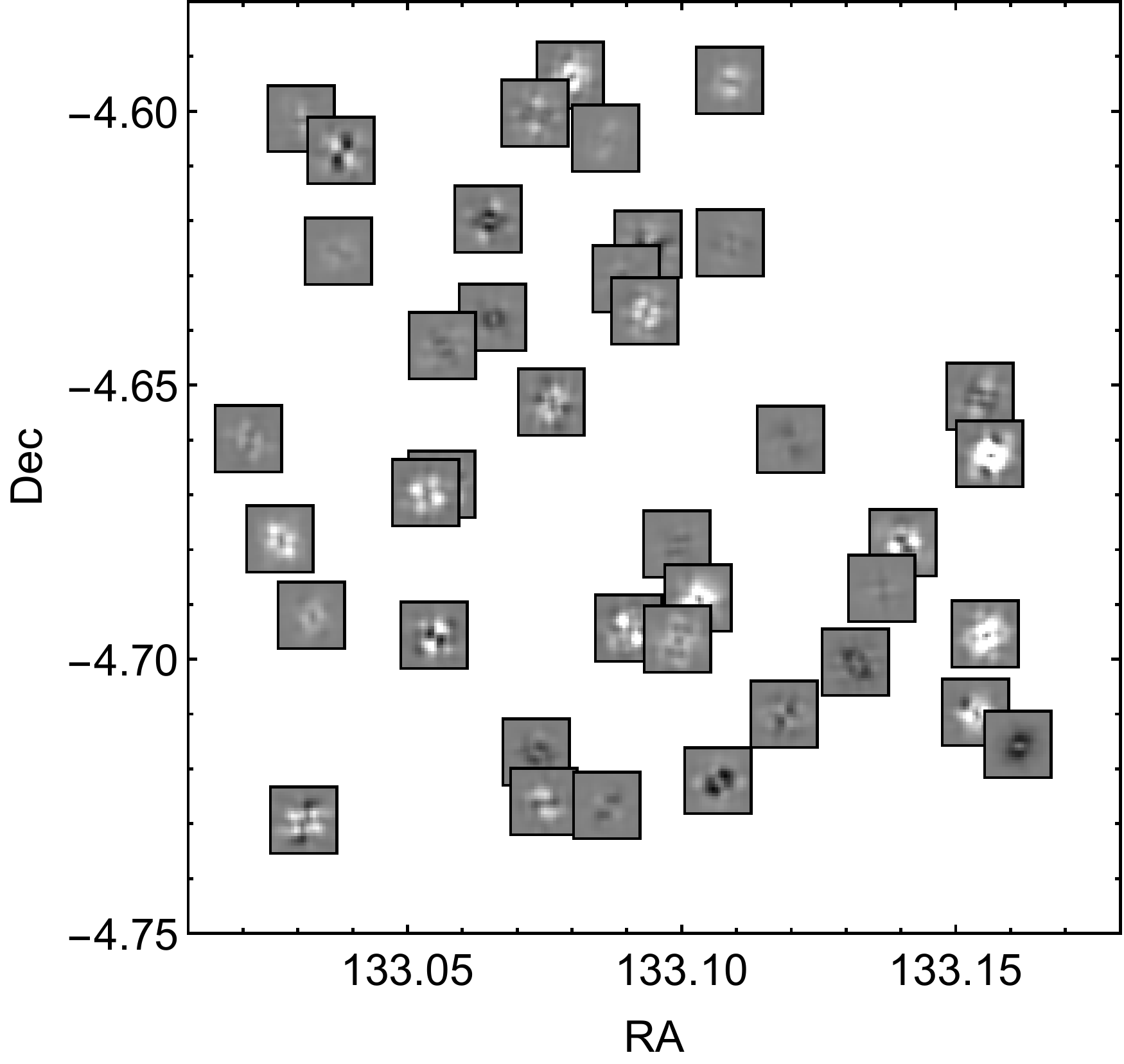}
  \caption{The distribution of the residual PSF power spectra in a small region of the w2m0m0 field of CFHTLenS. The PSF model is constructed with the 1st-order chipwise polynomial fitting. The residuals are calculated at the positions of the stars in the ``reference'' group.}
  \label{fig:psf-residual}
\end{figure}

As an example, we show in Fig.\ref{fig:psf-residual} the spatial distribution of a number of residual PSF power spectra derived from an exposure of the w2m0m0 field of the CFHTLenS data.

PCA provides a way of characterizing/parameterizing the morphologies of the residual PSF power spectra. It converts a set of correlated random variables into a set of linearly uncorrelated variables such that their variances are sorted in descending order. Regarding image processing for our purpose, PCA can be done by eigenvalue decomposition of the covariance matrix, which is defined as: 
\begin{equation}
\mathbf{A}_{ij}=\frac{1}{N-1}\sum_{m=1}^{N}{\left(\bfr^{[m]}_i-\mu_i\right)\left(\bfr^{[m]}_j-\mu_j\right)},
\end{equation}
where $\bfr^{[m]}_i$ denotes the value of the $i^\mathrm{th}$ pixel ($i=1,2,\cdots,l^2$) in the $m^\mathrm{th}$ image, and $\mu_i$ is the average value of the $i^\mathrm{th}$ pixel in all images. Note that image here refers to the residual of the PSF power spectrum defined in Eqn.\eqref{eqn:res-def}, which has the dimension of $l\times l$. $N$ is the total number of PSF residuals used. The covariance matrix $\mathbf{A}$ is decomposed by:
\begin{equation}
\mathbf{A}=\mathbf{Q}\mathbf{\Lambda}\mathbf{Q}^{-1}, 
\end{equation}
where $\mathbf{\Lambda}=\mathrm{diag}(\lambda_1,\cdots,\lambda_{l^2})$ is the matrix of eigenvalues in descending order, and $\mathbf{Q}=[\epsilon_1,\cdots,\epsilon_{l^2}]$ is the matrix of normalized eigenvectors, \ie the Principal Components (PCs). 

\begin{figure}[!htb]
  \centering
  \includegraphics[width=8cm]{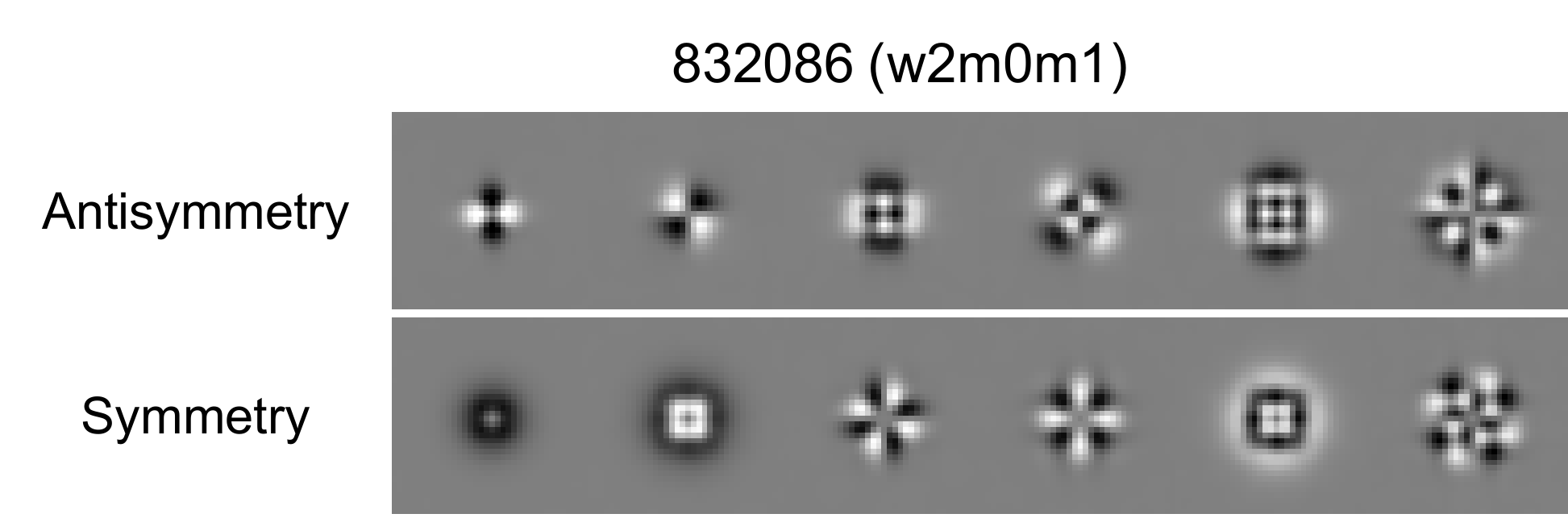} \\ \vspace{4pt}
  \includegraphics[width=8cm]{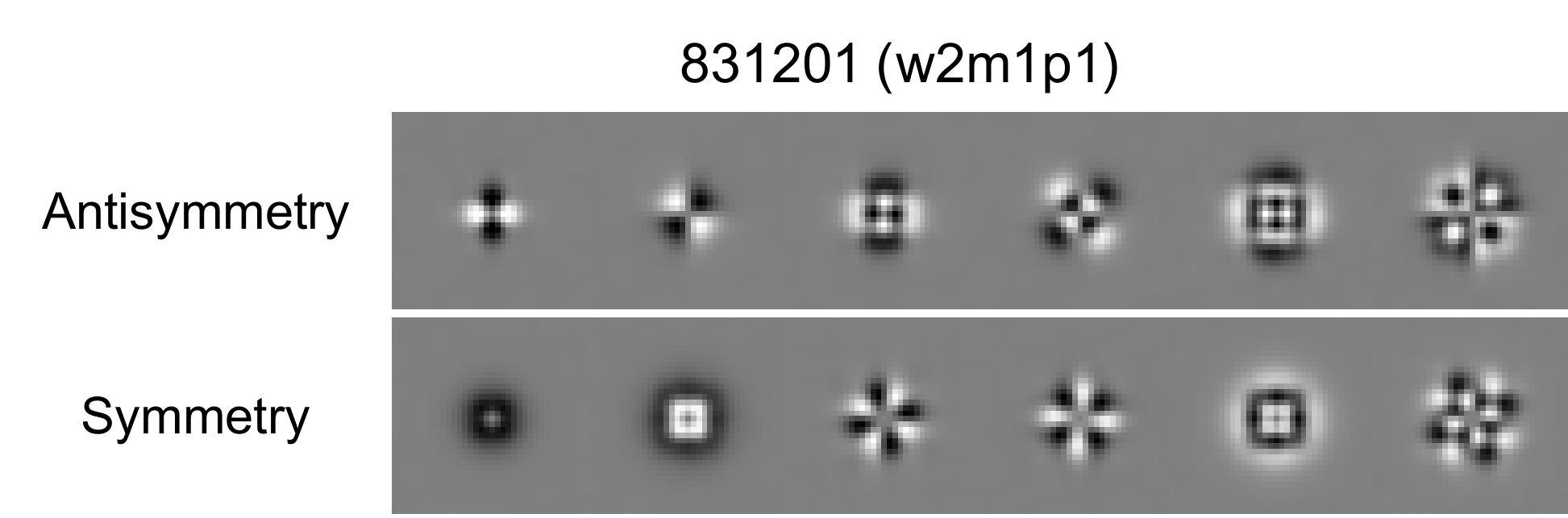}
  \caption{The first 6 principal components with $90^\circ$ antisymmetry (upper panel) and the 6 principal components with $90^\circ$ symmetry (lower panel) of PSF residuals on CFHTLenS 832086 (w2m0m1) and 831201 (w2m1p1).}
  \label{fig:pc}
\end{figure}

\begin{figure}[!htb]
  \centering
  \includegraphics[width=8.5cm]{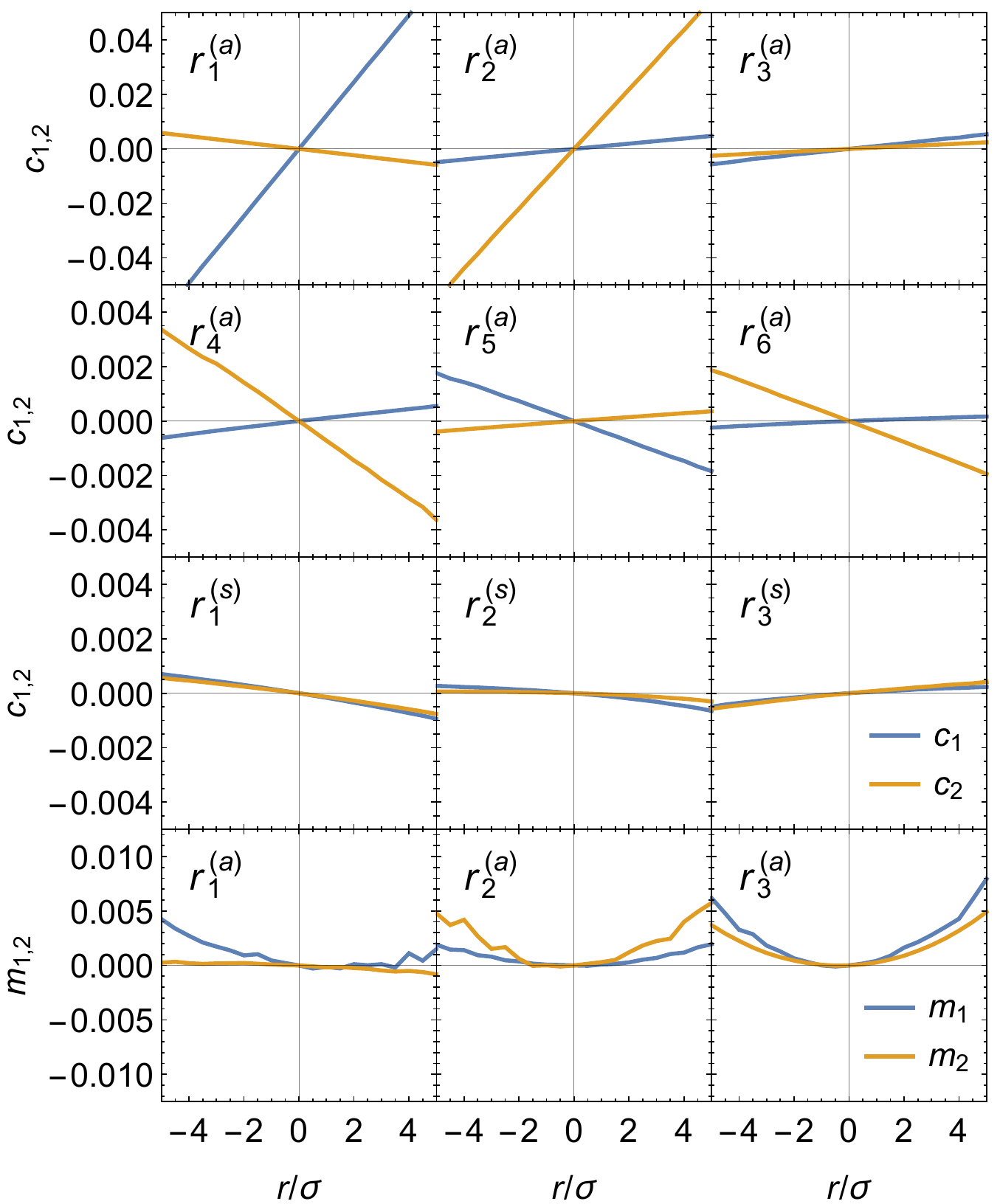}
  \caption{The relation between the residuals in both types of principal components and the induced biases are shown for 832086 (w2m0m1). The first two rows show the relations between antisymmetrical PCs and additive biases. The third row shows the relations between symmetrical PCs and additive biases. The last row shows the relation between antisymmetrical PCs and multiplicative biases. The residuals $r_p$ are divided by their corresponding standard deviations $\sigma_p$. In each of these panels, a $5\sigma$ range of the residual is shown. Galaxies are generated by 20 point sources following two-dimensional normal distribution with $\sigma=2.1\,\mathrm{pixel}$.}
  \label{fig:cvi}
\end{figure}

Our PCA is performed using only the stars on the same exposure. There are typically of order ten PCs that can have significant eigenvalues. However, not all of the PCs are relevant for the additive shear biases. One way to reduce the degree of freedom is to decompose the PSF residual power into two parts -- a $90^\circ$ symmetry part $\bfr^\mathrm{(s)}$ and a $90^\circ$ antisymmetry part $\bfr^\mathrm{(a)}$: 
\begin{eqnarray}
\nonumber \bfr^\mathrm{(s)}(k_x,k_y)&:=&\frac{1}{2}\left[\bfr(k_x,k_y)+\bfr(-k_y,k_x)\right], \\
\bfr^\mathrm{(a)}(k_x,k_y)&:=&\frac{1}{2}\left[\bfr(k_x,k_y)-\bfr(-k_y,k_x)\right].
\end{eqnarray}
PCA can be performed on $\bfr^\mathrm{(s)}$ and $\bfr^\mathrm{(a)}$ individually to get $\mathbf{Q}^\mathrm{(s)}$ and $\mathbf{Q}^\mathrm{(a)}$. Fig.\ref{fig:pc} shows the examples of these two types of PCs using the data on two different exposures of CFHTLenS. We find that due to the symmetry property of cosmic shear, the $90^\circ$ symmetry part are not relevant to additive shear biases. One can check this statement by studying the responses of the additive shear biases to the PCs through Eqn.\eqref{eqn:bias-expansion}. This can be done by adding a specific PC to a true PSF power spectrum and plotting the additive biases against the amplitude of that PC. For this purpose, we define the principal coefficients $r_p^\mathrm{(s)}$ and $r_p^\mathrm{(a)}$ as:
\begin{eqnarray}
\bfr^\mathrm{(s)}&=&\sum_{p=1}^{l^2}{r_p^\mathrm{(s)} \epsilon_p^\mathrm{(s)}}, \\
\bfr^\mathrm{(a)}&=&\sum_{p=1}^{l^2}{r_p^\mathrm{(a)} \epsilon_p^\mathrm{(a)}}.
\end{eqnarray}
As is shown in Fig.\ref{fig:cvi}, the responses of additive biases to the principal coefficients of $90^\circ$ antisymmetrical PCs are within linear ranges, and $90^\circ$ symmetrical PCs have much lower impacts on the biases compared with the antisymmetrical PCs. Because our goal is to remove correlations between additive biases, we only consider the $90^\circ$ antisymmetrical PCs in this work, and omit the upper indices $(\mathrm{s})$ and $(\mathrm{a})$ hereafter. We also notice, in Fig.\ref{fig:cvi}, that the impacts of the coefficients of antisymmetrical PCs on multiplicative biases are negligible ($m_{1,2}\ll 1$). This result indicates the fact that when we modify PSFs on their antisymmetrical PCs, we will not change the values of multiplicative biases by much, and their correlation functions are expected to stay the same.

\begin{figure}[!htb]
  \centering
  \includegraphics[width=8.5cm]{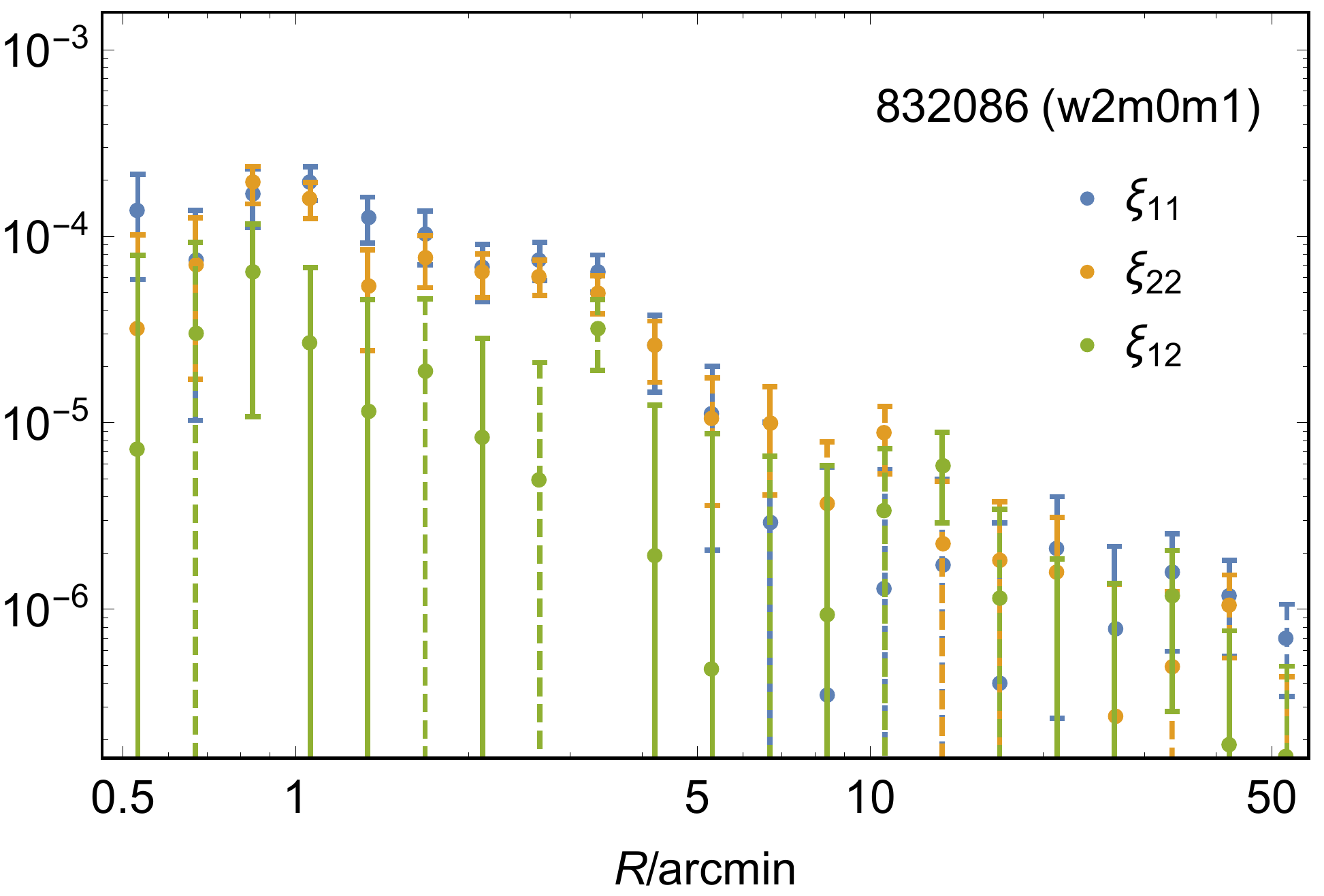} \\ \vspace{10pt}
  \includegraphics[width=8.5cm]{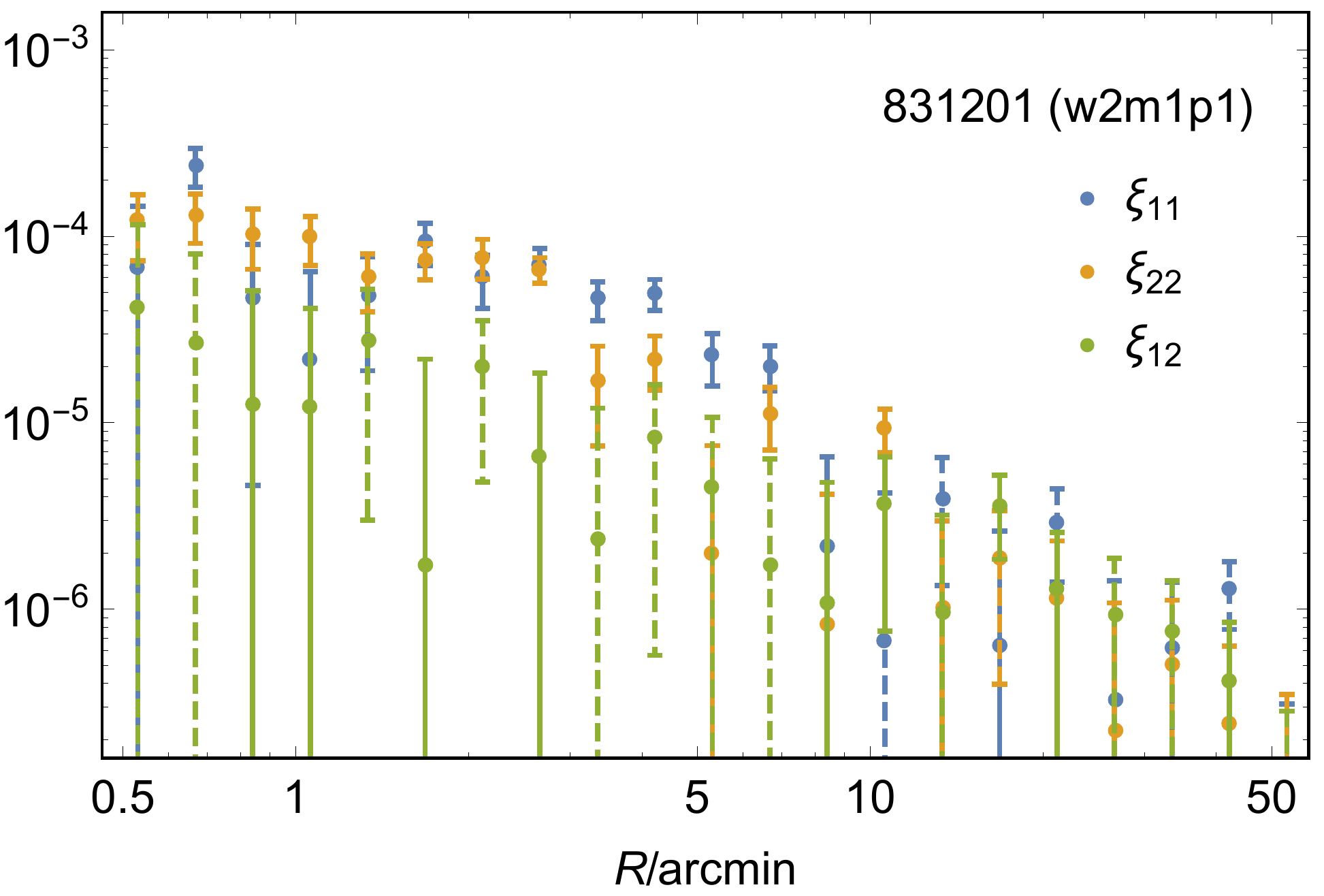}
  \caption{Auto- and cross-correlation functions between the first two components for two exposures. A dashed error bar means the value is negative.}
  \label{fig:corr-12}
\end{figure}

According to Eqn.\eqref{eqn:bias-expansion}, the effect of PSF residuals on the shear-shear correlations can be related to the spatial correlation functions between the $p$-th and the $q$-th PCs. In this paper, we only focus on measuring and removing the isotropic parts of correlation functions, which are defined as:
\begin{equation}
\xi_{pq}(R)=\langle r_p(\bfx)r_q(\bfx+R)\rangle.
\end{equation}
Here $\bfx+R$ means $\bfx+\bfy$ for all $|\bfy|=R$. Fig.\ref{fig:corr-12} shows some of the measured correlation functions of PCs. We find that the cross-correlations between the coefficients of different PCs are smaller than their auto-correlations. It turns out that the quantification of these correlation functions of the PCs provide us a direct way of correcting for the residual PSF uncertainties on shear-shear correlations, as shown in the next section. 

\subsection{Two PSF Fields, Two Shear Catalogs, and Bias-Free Shear-Shear Correlations}
\label{sec:corr-field}

The spatial correlations of the residual PSF power spectra are responsible for the biases of the shear-shear correlations. To correct for this effect, our basic idea is to introduce additional stochastic components (PCs) to the PSF field so that the spatial correlations between the additional components can cancel out that of the original PSF residuals. Technically, we find it more convenient to follow a slightly different approach: generating two PSF fields with additional anticorrelated stochastic components. The resulting two shear catalogs can be cross-correlated to yield a bias-free estimation of the shear-shear correlation function. For this purpose, we generate a set of Gaussian random fields $w_{pq}$ by doing two consecutive Fourier transformations on target correlation functions $\xi_{pq}$, defined as $\xi_{pq}(\bfx):=\xi_{pq}(R=|\bfx|)$: 
\begin{eqnarray}
S_{pq}(\boldsymbol\omega)&=&\int{\xi_{pq}(\bfx) e^{-i\boldsymbol\omega\cdot\bfx}\dd\bfx}, \label{eqn:power-spectrum}\\
D_{pq}(\boldsymbol\omega)&=&N(0,1)\sqrt{S_{pq}(\boldsymbol\omega)},\\
w_{pq}(\bfx)&=&\mathrm{Re}\int{D_{pq}(\boldsymbol\omega) e^{i\boldsymbol\omega\cdot\bfx}\dd\boldsymbol\omega},
\end{eqnarray}
where $S_{pq}$ is the Fourier transform of $\xi_{pq}$, \ie the co-spectrum of $r_p$ and $r_q$, $D_{pq}$ is the randomized spectral density, $N(0,1)$ is a complex zero-mean unit-variance Gaussian random variable, and $w_{pq}$ is the desired Gaussian random field. According to Wiener--Khinchin theorem (\cite{wiener1930}; \cite{khinchin1934}), we have: 
\begin{eqnarray}
\nonumber\langle w_{pq}(\bfx)w_{pq}(\bfx+R)\rangle
&=&\langle w_{pq}(\bfx)w_{pq}(\bfx+\bfy)\rangle_{|\bfy|=R} \\
\nonumber&=&\langle \xi_{pq}(\bfy)\rangle_{|\bfy|=R} \\
&=&\xi_{pq}(R).
\end{eqnarray}
Note that since each $w_{pq}(\bfx)$ is generated independently, we have: 
\begin{equation}
\langle w_{pk}(\bfx)w_{jq}(\bfx+R)\rangle=\delta_{pj}\delta_{kq}\xi_{pq}(R).
\end{equation}

\begin{figure}[!htb]
  \centering
  \includegraphics[width=8cm]{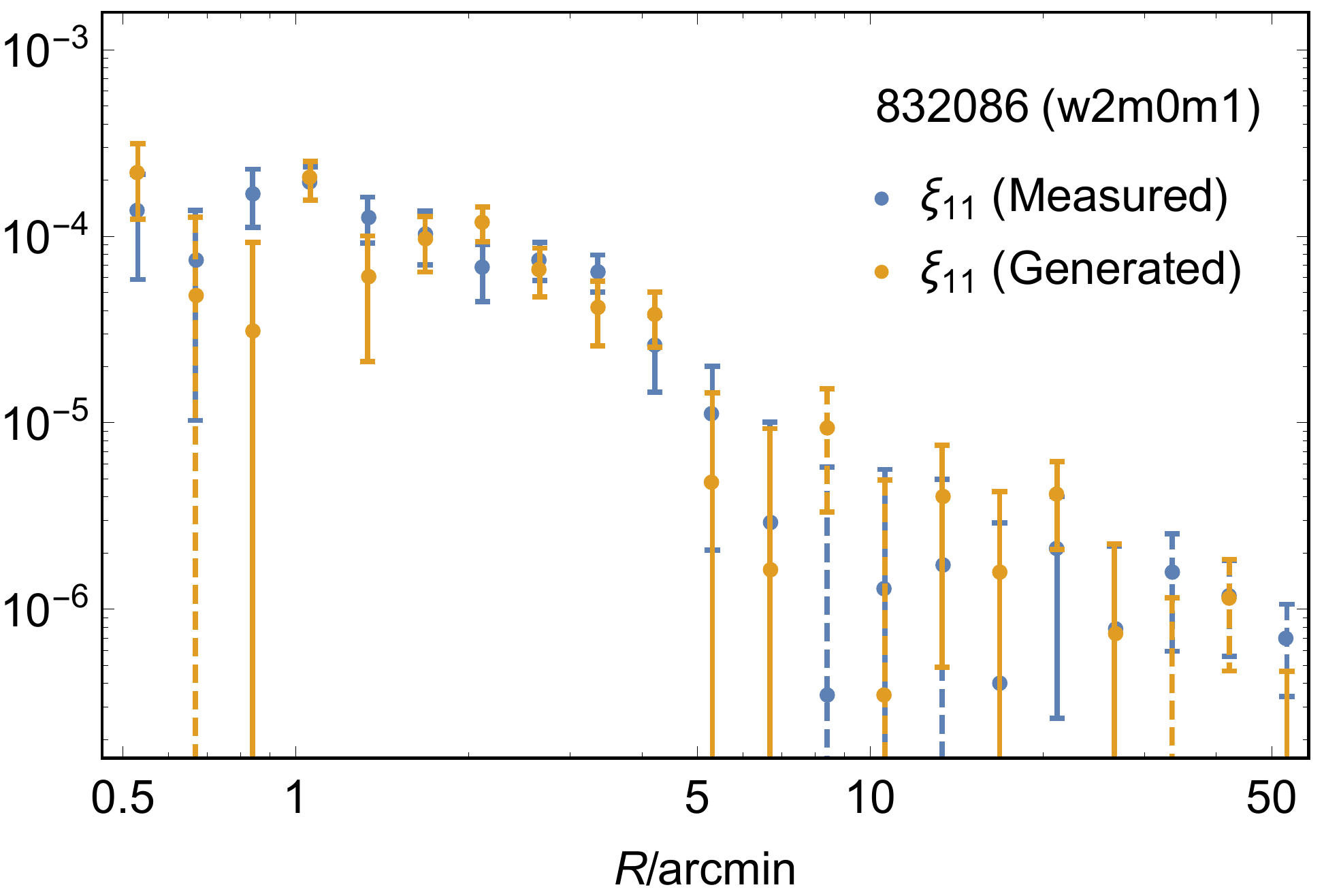} \\ \vspace{10pt}
  \includegraphics[width=8cm]{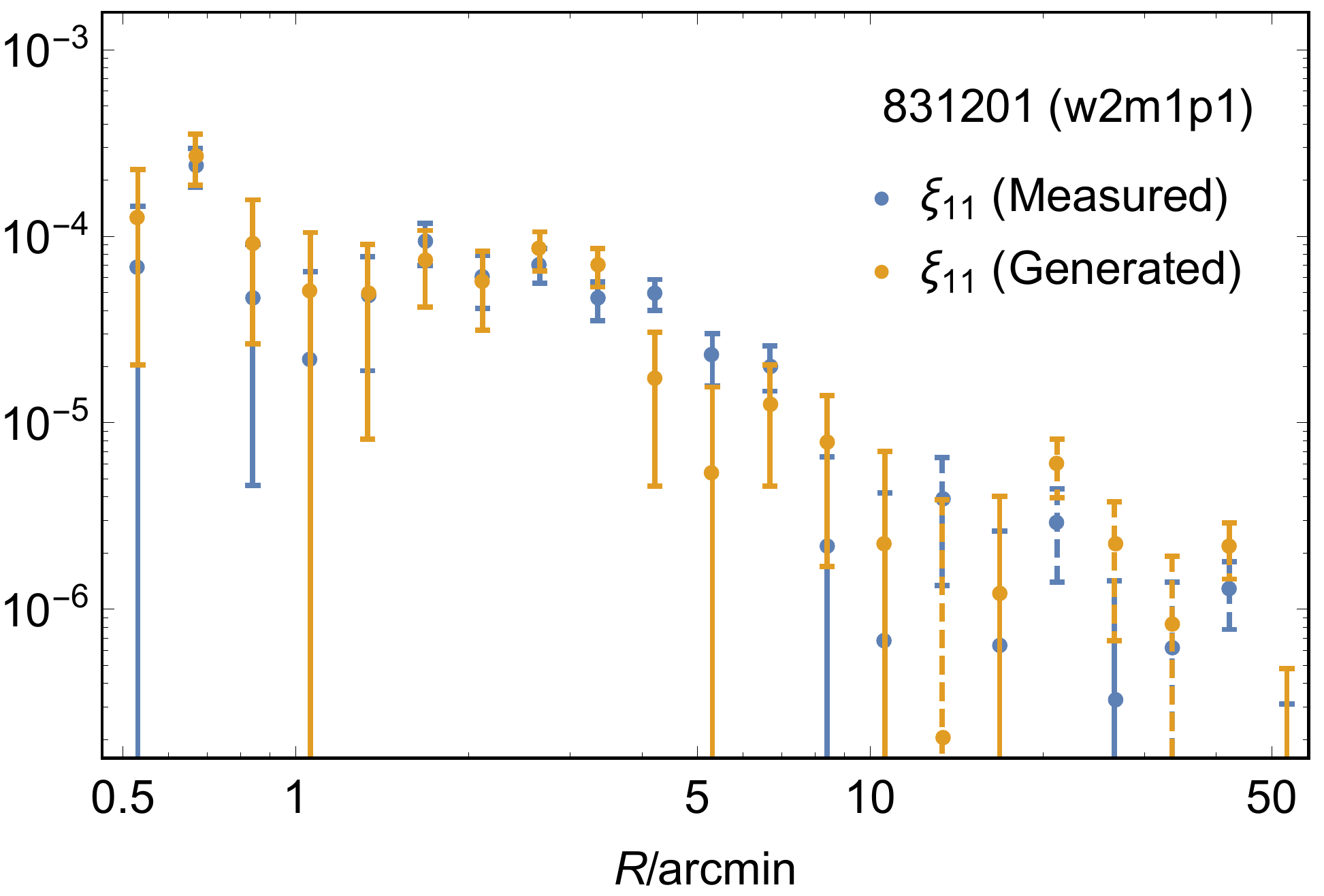}
  \caption{$\xi_{11}$ measured from real data and generated Gaussian random fields for two exposures. The correlation functions of the generated fields are averaged between 16 generations each.}
  \label{fig:corr-gen}
\end{figure}

Since the co-spectrum $S_{pq}$ serves as the power spectrum of $w_{pq}(\bfx)$, it is valid only if it is non-negative for all $\boldsymbol\omega$. This requirement on $S_{pq}$ is often not satisfied because co-spectrums between different fields can be negative for some $\boldsymbol\omega$, and those of identical fields can also have negative power values due to inaccurate measurements of $\xi_{pq}(R)$. However, we notice that the correlation functions at very short range ($\lsim R_0=0.5\,\mathrm{arcmin}$) are not relavant to weak lensing measurements. Therefore, we can adjust this part of each correlation function to make $S_{pq}$ a valid power spectrum for $w_{pq}(\bfx)$. One of the approaches is to increase the value of $\xi_{pq}(0)$ only, but it can easily invalidate Eqn.\eqref{eqn:bias-expansion} due to a large increase in the variances of generated fields, \ie $\xi_{pq}(0)^2$. The approach we take is to minimize the variances of generated fields during adjusting to maintain the accuracy of Eqn.\eqref{eqn:bias-expansion}. According to Eqn.\eqref{eqn:power-spectrum}, this adjusting process can be described as an optimization problem\footnote{After discretizations on $\bfx$ and $\boldsymbol\omega$ spaces, this problem is transformed into a linear programming problem on $\xi_{pq}(\bfx)$ with $|\bfx|<R_0$, which can be solved much faster. We project $\xi_{pq}(R)$ to its two-dimensional counterpart $\xi_{pq}(\bfx)$ using nearest-neighbor interpolation.}:
\begin{eqnarray}
\nonumber
&&\argmin_{\xi_{pq}(\bfx)\,(|\bfx|<R_0)}{\xi_{pq}(0)^2},\\
&&\textrm{subject to}\;\;\forall\;\boldsymbol\omega:\int{\xi_{pq}(\bfx) e^{-i\boldsymbol\omega\cdot\bfx}\dd\bfx}>0.
\end{eqnarray}
As is shown in Fig.\ref{fig:corr-gen}, the correlation function of generated field is the same as the target correlation function within a margin of error.

Based on the $w_{pq}$ fields and the PCs $\epsilon_p$, we can build up two PSF fields to modify the original PSF model:
\begin{eqnarray}
\bfu^+(\bfx)&=&\bfu(\bfx)+\bfr^+(\bfx) \\
\bfu^-(\bfx)&=&\bfu(\bfx)+\bfr^-(\bfx).
\end{eqnarray}
where
\begin{eqnarray}
\bfr^{+}(\bfx)&=&\sum_{p}{r_p^+ \epsilon_p}:=\sum_{p,q}{\sqrt{\frac{\sigma_p}{\sigma_q}}\,w_{pq}(\bfx)\epsilon_p},\nonumber \\
\bfr^{-}(\bfx)&=&\sum_{q}{r_q^- \epsilon_q}:=-\sum_{q,p}{\sqrt{\frac{\sigma_q}{\sigma_p}}\,w_{pq}(\bfx)\epsilon_q}.
\label{eqn:r-p-def}
\end{eqnarray}
$\sigma_p$ is the standard deviation of the $p$-th principal coefficients, 
and $\sqrt{\sigma_p/\sigma_q}$ are called the normalization factors. Note that each Gaussian random field $w_{pq}$ will impact the $p$-th and $q$-th principal coefficients at the same time, and the normalization factors balance these impacts based on their original standard deviations, ensuring the accuracy of Eqn.\eqref{eqn:bias-expansion}. 

The two PSF fields lead to two shear catalogs $\{\tilde{g}^+_1,\tilde{g}^+_2\}$ and $\{\tilde{g}^-_1,\tilde{g}^-_2\}$. If we let the calculation of the shear-shear correlation function to be defined by the cross-correlation between the two shear catalogs, we find that the contamination from the correlated PSF residuals automatically disappears. For example, the new shear-shear correlation $\xi_+(r)$ is defined as:
\begin{equation}
\tilde{\xi}_+(R):=\langle\tilde{g}^+_1(\bfx)\tilde{g}^-_1(\bfx+R)+\tilde{g}^+_2(\bfx)\tilde{g}^-_2(\bfx+R)\rangle.
\label{eqn:new-def}
\end{equation}
In the appendix, we show that $\tilde{\xi}_+(R)$ indeed leads to a unbiased measurement of the shear-shear correlation function.

\section{Results}
\label{sec:results}

We use the CFHTLenS image data (Erben et al. 2013) to illustrate our method. We download the Elixir-preprocessed i'-band single exposures of the W2 field from the Canadian Astronomical Data Center (CADC), as W2 has the highest stellar number density among the four fields. The basic preprocessing (background smoothing, cosmic-ray identification, astrometric correction, etc.) of the exposures are conducted using the THELI software developed by the CFHTLenS team (Erben et al. 2005; Schirmer 2013). The stars and galaxies are separated using the conventional ’Radius-Magnitude’ plot. We only retain stars with signal-to-noise-ratio (SNR) larger than 100 for avoiding ambiguities. The star images, as well as the mock galaxy images, are contained in $48\times 48$ stamps. We run additional routines to remove certain problematic images, such as those contaminated by bad pixels, cosmic rays, neighboring sources (binary stars), etc.. As a result, each chip in W2 contains about 100 stars. The results in this work uses 104 exposures from 16 pointings of the W2 field. More details regarding our processing of the CFHTLenS data can be found in L17. 

\begin{figure}[!htb]
  \centering
  \includegraphics[width=8cm]{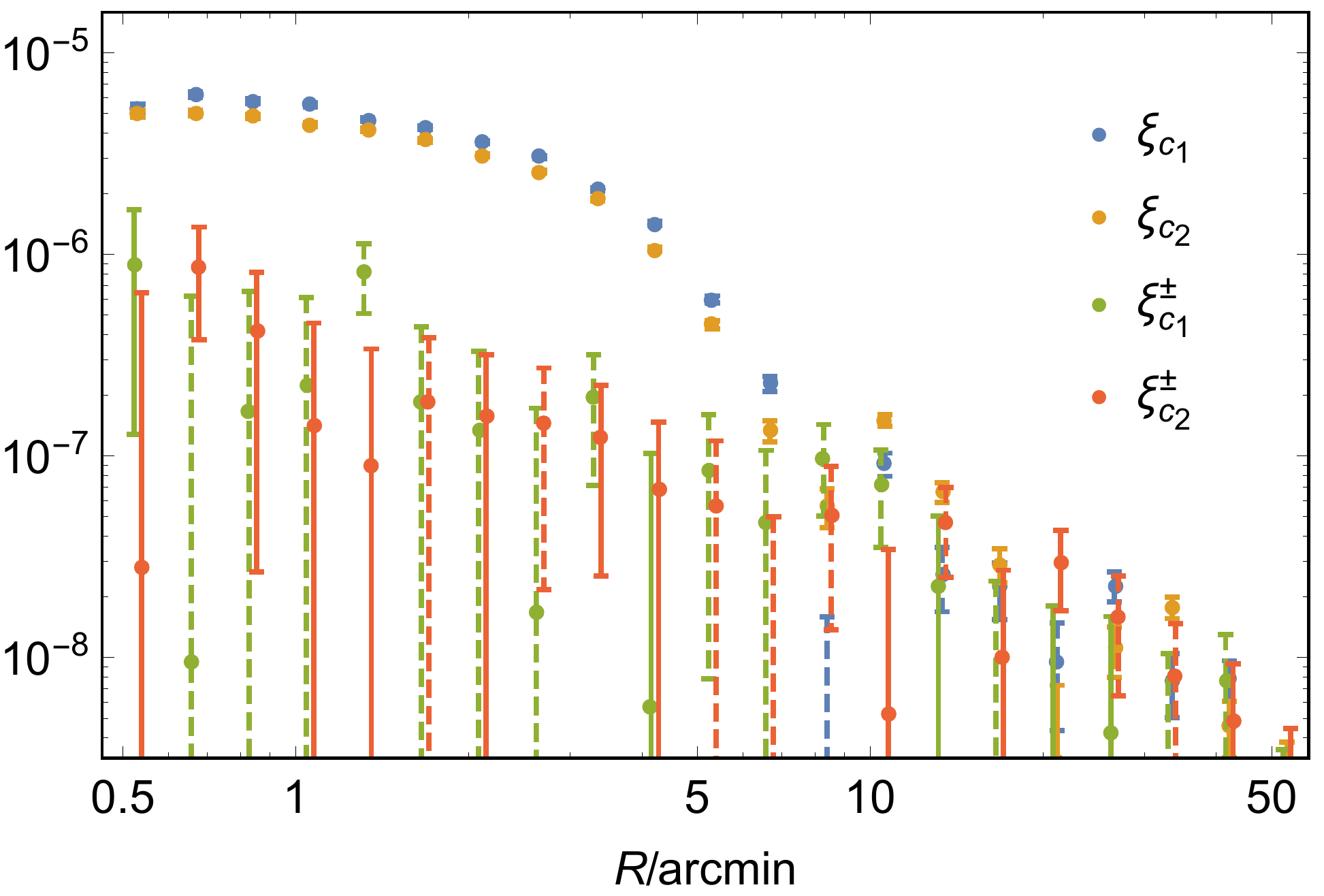}\\ \vspace{10pt}
  \includegraphics[width=8cm]{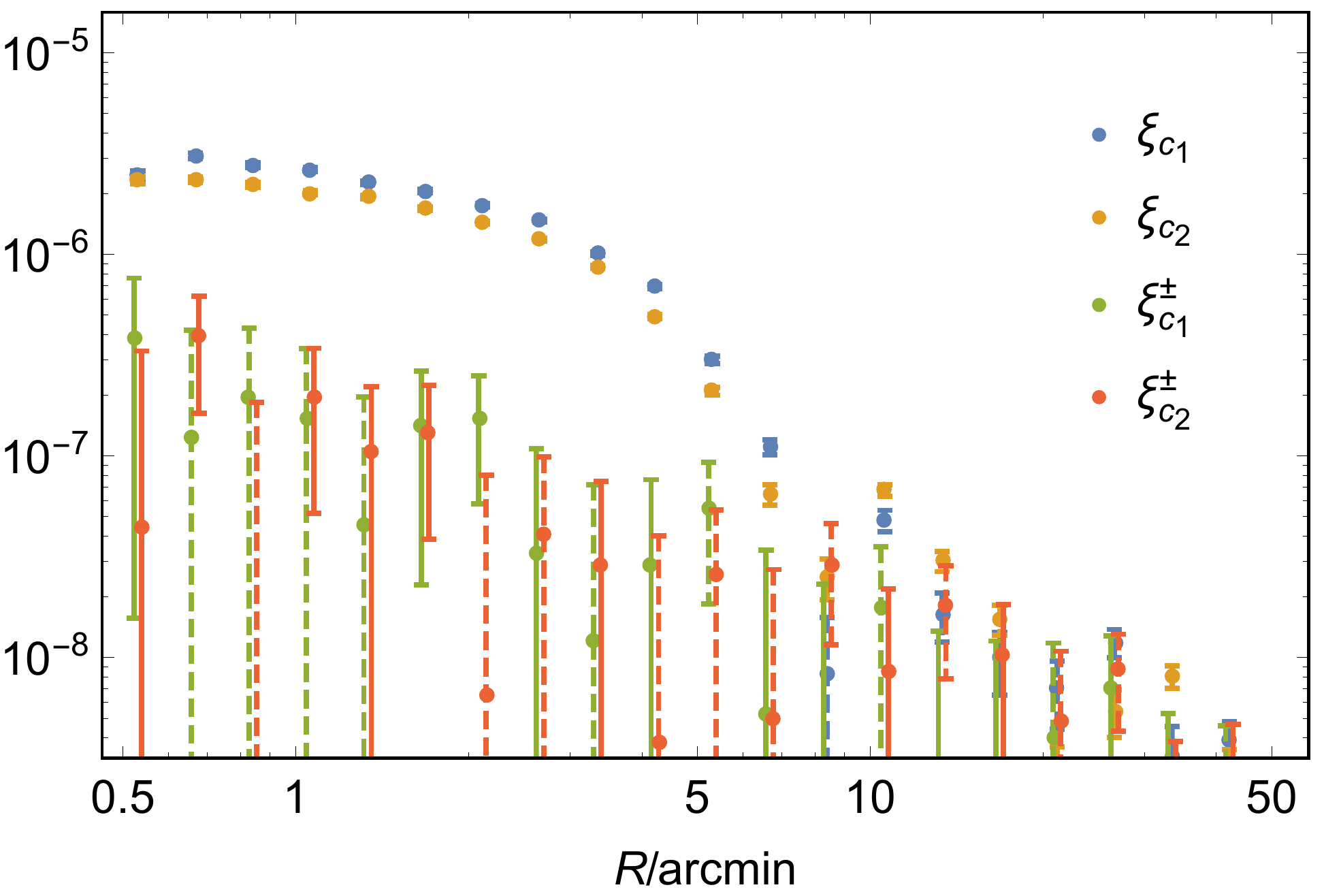}\\ \vspace{10pt}
  \includegraphics[width=8cm]{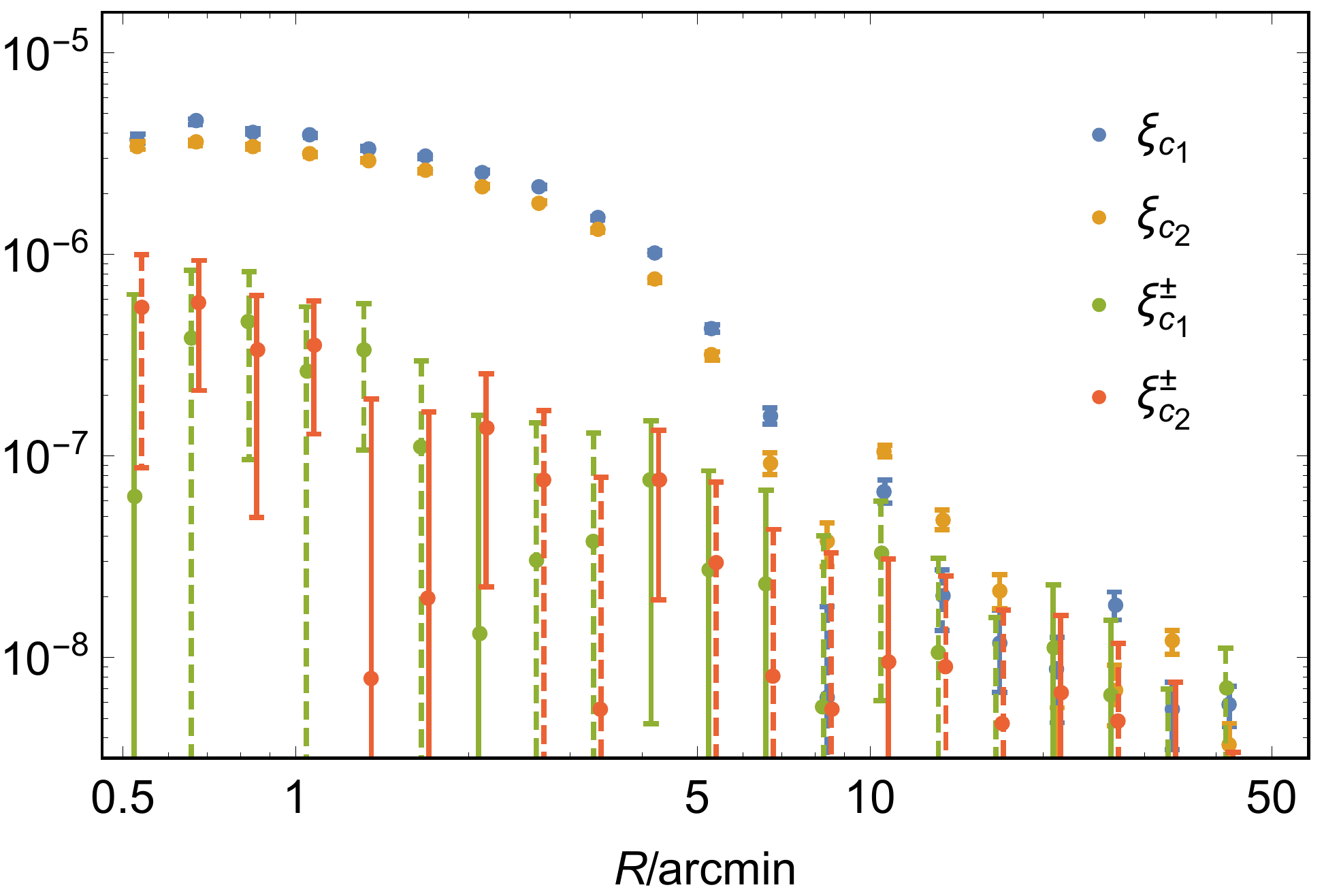}
  \caption{Residual additive bias correlations of three groups of galaxies: 1) $\sigma=2.1$ (upper), 2) $\sigma=3.5$ (middle), and 3) a mixture of $\sigma=2.1$ and $\sigma=3.5$ (lower). The results are measured based on CFHTLenS w2, including 16 pointings and 7 exposures on each pointing. The data points are offset by a small distance to improve visibility, and a dashed error bar means the value is negative.}
  \label{fig:cfvc}
\end{figure}

\begin{figure}[!htb]
  \centering
  \includegraphics[width=8cm]{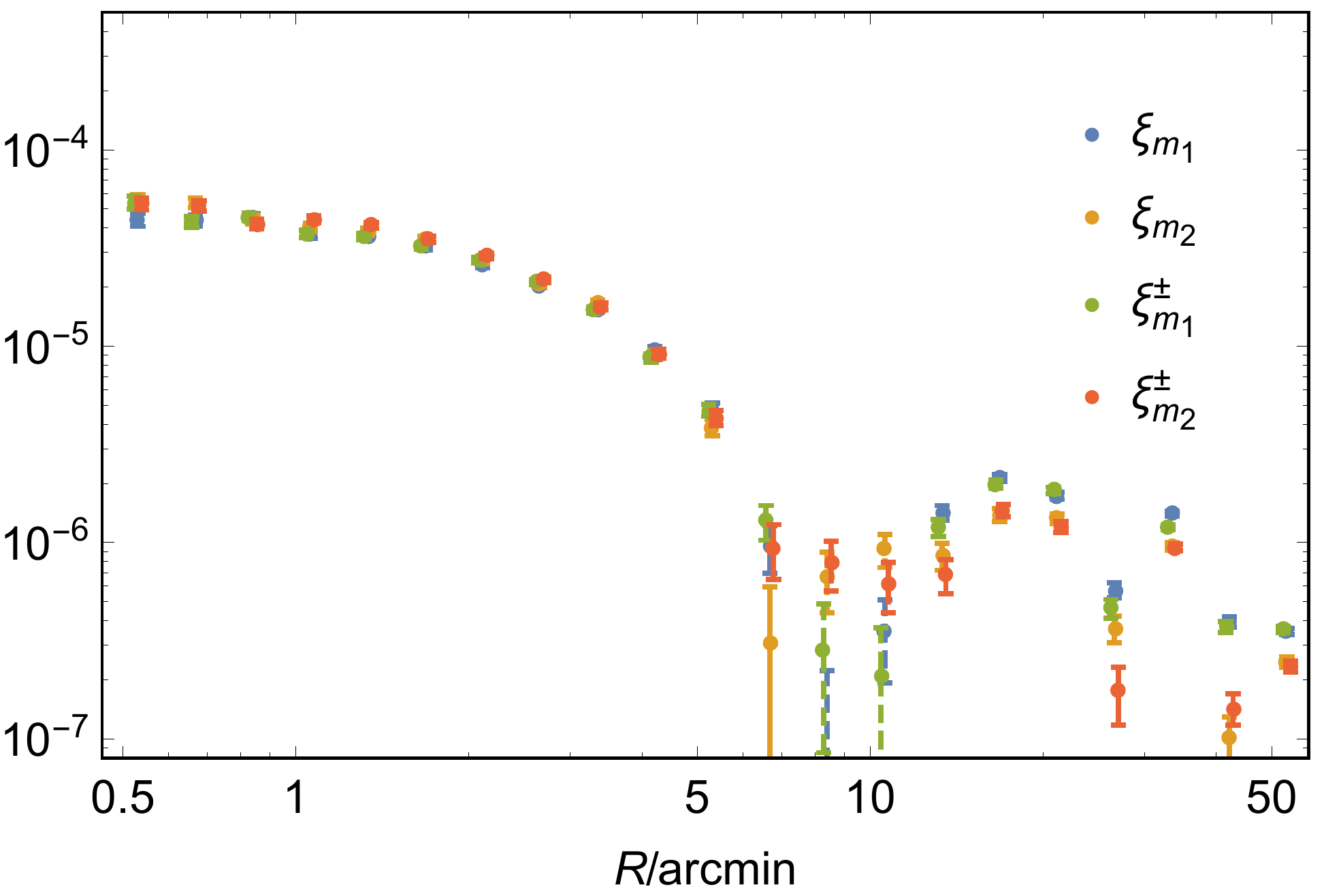}
  \caption{Residual multiplicative bias correlations with galaxy size $\sigma=2.1$. The correlation function of multiplicative biases are defined similar to those of additive biases in Eqn.\eqref{eqn:bias-corr-def1} and Eqn.\eqref{eqn:bias-corr-def2}. Other details are the same as Fig.\ref{fig:cfvc}.}
  \label{fig:cfvm}
\end{figure}

We illustrate the effect of correlation removals by comparing it with the original additive bias correlations, which directly show the errors in shear correlations given the morphology of galaxies. Without correlation removals, the bias correlations for $g_1$ and $g_2$ are:
\begin{equation}
\xi_{c_i}(R)=\langle c_i(\bfx)c_i(\bfx+R)\rangle \hspace{10pt} (i=1,2).
\label{eqn:bias-corr-def1}
\end{equation}
While with correlation removals, the residual correlations become: 
\begin{equation}
\xi^\pm_{c_i}(R)=\langle c^+_i(\bfx)c^-_i(\bfx+R)\rangle \hspace{10pt} (i=1,2).
\label{eqn:bias-corr-def2}
\end{equation}

We calculate these correlation functions with a large number of mock galaxies assigned at every validation group star position. Each of our mock galaxies is made of 20 point sources of equal luminosities, with a 2D Gaussian distribution. We directly generate the power spectrum of the galaxy image using the PSF form given by the validation group star. Note that each point source simply contributes a plane-wave in Fourier space. The power spectrum of the galaxy and the reconstructed PSF are used together for shear recovery with the method of ZLF15, and the additive shear biases are evaluated. We test our method for three groups of galaxies:
\begin{enumerate}
\item galaxies with sizes near $\sigma=2.1\,\mathrm{pixel}$;
\item galaxies with sizes near $\sigma=3.5\,\mathrm{pixel}$;
\item half of galaxies with sizes near $\sigma=2.1\,\mathrm{pixel}$ and half near $\sigma=3.5\,\mathrm{pixel}$.
\end{enumerate}

Our method induces additional stochasticity to individual PSF fields, causing large variances in the results. To reduce the variances, we generate 40 random Gaussian fields for each $\xi_{pq}$, which is equivalent to use each exposure 16 times. This technique can be used in real measurements in the exact same way. 

The results are shown in Fig.\ref{fig:cfvc}. The residual correlations are reduced by a factor of 10 in mid-range ($R\sim3\,\mathrm{arcmin}$). The performance of the mixed group lies between those of two groups with single sizes. This fact indicates that our method is not affected by the variety of galaxies. We also show a result for residual multiplicative biases correlations in Fig.\ref{fig:cfvm}. We find that our method does not change these correlation functions, as is expected due to negligible responses of multiplicative biases to $90^\circ$ antisymmetrical PCs.

\section{Conclusion}
\label{sec:conclusion}

Current PSF reconstruction methods are not able to recover the fluctuations of the PSF morphologies on small scales. The spatial correlations of the residual PSF uncertainties are significant contaminations to the shear-shear correlations. To overcome this problem, we study the properties of the residual PSF field with the Principal-Component-Analysis. The coefficients of the PCs of the residual PSF images are found to be spatially correlated, and are linearly related to the additive shear bias. Based on these facts, we propose a way to achieve unbiased shear-shear correlations: first, we create two PSF fields by adding two different sets of additional stochastic and correlated PCs to the PSF model; secondly, two shear catalogs are generated by the two sets of PSF fields respectively; finally, the shear-shear correlations are measured by cross-correlating the shears from the two different catalogs. Our method is tested using the imaging data from the W2 field of CFHTLenS, and the shear measurement method of ZLF15. The PSF model is constructed using stars of SNR $> 100$, and the first-order chipwise polynomial fitting. Our numerical results show that the bias to the shear-shear correlations can be significantly reduced at the angular range of a few arcmins. 


In principle, the idea of this paper also works for other shear measurement methods. It only requires one to identify a number of PCs that can significantly influence the additive shear biases through linear relations, such as our Eqn.\eqref{eqn:bias-expansion}. However, certain details that distinguishes ZLF15 from other shear measurement methods can affect the quanlity of the final results. For example, unlike most other shear measurement methods, ZLF15 only requires the PSF power spectrum, so that there are no alignment issues in PSF reconstruction and PCA. The validity of the linear relations between the PC coefficients of the residual PSF and the additive shear biases should also be studied specifically for each shear measurement method. These issues will be discussed in a future work.

\acknowledgments{This work is based on observations obtained with MegaPrime/MegaCam, a joint project of CFHT and CEA/DAPNIA, at the Canada-France-Hawaii Telescope (CFHT) which is operated by the National Research Council (NRC) of Canada, the Institut National des Sciences de l\'Univers of the Centre National de la Recherche Scientifique (CNRS) of France, and the University of Hawaii. This research used the facilities of the Canadian Astronomy Data Centre operated by the National Research Council of Canada with the support of the Canadian Space Agency. CFHTLenS data processing was made possible thanks to significant computing support from the NSERC Research Tools and Instruments grant program.

The processing of single exposure images is conducted using the THELI software, a tool for the automated reduction of astronomical images developed by the CFHTlenS team \citep{erben2005,schirmer2013}.

JZ is supported by the NSFC grants (11673016, 11433001, 11621303) and the National Key Basic Research Program
of China (2015CB857001). L.P.F. acknowledges the
support from NSFC grant 11722326, 11673018 \& 11333001, STCSM grants 16R1424800 and SHNU grant DYL201603. D.Z.Liu and Z.H.Fan are supported in part by NSFC of China under the grants 11333001 and 11653001.

}

\appendix
\section{Bias-Free Correlation Function}
Regarding the correlation function $\tilde{\xi}_+(R)$ defined in Eqn.\eqref{eqn:new-def}, we show that its expectation value is not affected by the existence of the correlated additive shear biases due to the PSF residuals. For convenience, we only consider the additive biases. The proof is shown below, with $|\bfx-\bfy|=R$:

\begin{align*}
\tilde{\xi}_+(R)&=\left\langle\tilde{g}^+_1(\bfx)\tilde{g}^-_1(\bfy)\right\rangle+\left\langle\tilde{g}^+_2(\bfx)\tilde{g}^-_2(\bfy)\right\rangle \\
&=\left\langle\left(g_1(\bfx)+c^+_1(\bfx)\right)\left(g_1(\bfy)+c^-_1(\bfy)\right)\right\rangle+[\textrm{terms for }g_2] \\
&=\left\langle\left(g_1(\bfx)+\sum_p{\alpha_{1p}(\bfx)(r_p(\bfx)+r^+_p(\bfx))}\right)\left(g_1(\bfy)+\sum_q{\alpha_{1q}(\bfy)(r_q(\bfy)-r^-_q(\bfy))}\right)\right\rangle+[\textrm{terms for }g_2] \\
&=\left\langle g_1(\bfx)g_1(\bfy)\right\rangle+\sum_{p,q}{\left\langle\alpha_{1p}(\bfx)\alpha_{1q}(\bfy)\right\rangle\left\langle r_p(\bfx)r_q(\bfy)-r^+_p(\bfx)r^-_q(\bfy)\right\rangle}+[\textrm{terms for }g_2] \\
&=\left\langle g_1(\bfx)g_1(\bfy)\right\rangle+\sum_{p,q}{\left\langle\alpha_{1p}(\bfx)\alpha_{1q}(\bfy)\right\rangle\left\langle r_p(\bfx)r_q(\bfy)-\sum_j{\sqrt{\frac{\sigma_p}{\sigma_j}}\,w_{pj}(\bfx)}\sum_k{\sqrt{\frac{\sigma_q}{\sigma_k}}\,w_{kq}(\bfy)}\right\rangle}+[\textrm{terms for }g_2] \\
&=\left\langle g_1(\bfx)g_1(\bfy)\right\rangle+\sum_{p,q}{\left\langle\alpha_{1p}(\bfx)\alpha_{1q}(\bfy)\right\rangle\left\langle r_p(\bfx)r_q(\bfy)-\sqrt{\frac{\sigma_p}{\sigma_q}}\,w_{pq}(\bfx)\sqrt{\frac{\sigma_q}{\sigma_p}}\,w_{pq}(\bfy)\right\rangle}+[\textrm{terms for }g_2] \\
&=\left\langle g_1(\bfx)g_1(\bfy)\right\rangle+\sum_{p,q}{\left\langle\alpha_{1p}(\bfx)\alpha_{1q}(\bfy)\right\rangle\left[\xi_{pq}(R)-\xi_{pq}(R)\right]}+[\textrm{terms for }g_2] \\
&=\left\langle g_1(\bfx)g_1(\bfy)+g_2(\bfx)g_2(\bfy)\right\rangle.
\end{align*}

\end{document}